\documentclass[prd, twocolumn, superscriptaddress, nofootinbib, notitlepage, floatfix, preprintnumbers]{revtex4-1}
\usepackage{xcolor}
\usepackage{graphicx}
\usepackage{wrapfig}
\usepackage{amsmath}
\usepackage{amsfonts}
\usepackage{amssymb}
\usepackage{multirow}
\usepackage{slashed}
\usepackage{physics}
\usepackage{soul}
\usepackage{ulem}
\usepackage[colorlinks=true, linkcolor=blue, citecolor=blue, urlcolor=blue]{hyperref}
\newcommand{\msbar}{{\overline{\mathrm{MS}}}}
\newcommand\befs{\begin{figure*}}
\newcommand\eefs[1]{\label{fig:#1}\end{figure*}}
\newcommand\bef{\begin{figure}}
\newcommand\eef[1]{\label{fig:#1}\end{figure}}
\newcommand\beq{\begin{equation}}
\newcommand\eeq[1]{\label{#1}\end{equation}}
\newcommand\beqa{\begin{eqnarray}}
\newcommand\eeqa[1]{\label{#1}\end{eqnarray}}
\newcommand\bet{\begin{table}}
\newcommand\eet[1]{\label{tb:#1}\end{table}}
\newcommand\bets{\begin{table*}}
\newcommand\eets[1]{\label{tb:#1}\end{table*}}
\newcommand{\argmax}[1]{\underset{#1}{\operatorname{arg}\,\operatorname{max}}\;}
\newcommand{\argmin}[1]{\underset{#1}{\operatorname{arg}\,\operatorname{min}}\;}
\newcommand\fgn[1]{Fig.\ \ref{fig:#1}}
\newcommand\eqn[1]{Eq.\ (\ref{#1})}

\newcommand{\om}{{\omega}}

\begin{document}
\widetext

\title{
    Bayesian-Wilson coefficients in lattice QCD computations of valence PDFs and GPDs
}
\author{Nikhil Karthik}
\email{nkarthik.work@gmail.com}
\affiliation{Department of Physics, College of William \& Mary, Williamsburg, VA 23185, USA}
\affiliation{Thomas Jefferson National Accelerator Facility, Newport News, VA 23606, USA}
\author{Raza Sabbir Sufian}
\email{sufian@jlab.org}
\affiliation{Department of Physics, College of William \& Mary, Williamsburg, VA 23185, USA}
\affiliation{Thomas Jefferson National Accelerator Facility, Newport News, VA 23606, USA}
\begin{abstract}
    We propose an analysis method for the leading-twist operator
    product expansion based lattice QCD determinations of the
    valence parton distribution function (PDF). In the first step, we determine
    the confidence-intervals of the leading-twist $\msbar$ Wilson
    coefficients, $C_n(\mu^2 z^2)$, of the equal-time bilocal quark
    bilinear, given the lattice QCD matrix element of Ioffe-time
    distribution for a particular hadron {\sl H} as well as the
    prior knowledge of the valence PDF, $f(x,\mu)$ of the hadron {\sl H}
    determined via global fit from the experimental data. In the
    next step, we apply the numerically estimated $C_n$ in the
    lattice QCD determinations 
    of the valence PDFs of other hadrons, and for
    the zero-skewness generalized parton distribution (GPD) of the
    same hadron {\sl H} at non-zero momentum transfers.  Our proposal
    still assumes the dominance of leading-twist terms, but it
    offers a pragmatic alternative to the usage of perturbative
    Wilson coefficients and their associated higher-loop uncertainties
    such as the effect of all-order logarithms at larger sub-Fermi
    quark-antiquark separations $z$. 
\end{abstract}

\preprint{JLAB-THY-21-3416}
\date{\today}
\maketitle

\section{Introduction}

The progress in determining the $x$-dependent hadron structures,
such as the parton distribution functions (PDFs) and the generalized
parton distribution functions
(GPDs)~\cite{Mueller:1998fv,Ji:1996ek,Radyushkin:1996nd} (for a
review, see~\cite{Diehl:2003ny}), has been quite rapid in the recent
years, particularly owing to the perturbative matching frameworks
such as the quasi-PDF~\cite{Ji:2013dva,Ji:2014gla},
pseudo-PDF~\cite{Radyushkin:2017cyf,Orginos:2017kos}, current-current
correlators~\cite{Braun:2007wv} and the good lattice cross-sections
approach~\cite{Ma:2014jla,Ma:2017pxb}.  There are other ideas that are
being used (for
example,~\cite{Martinelli:1987zd,Liu:1993cv,Chambers:2017dov,Detmold:2005gg,Detmold:2021uru}),
but we would restrict ourselves to the set of perturbative matching
approaches for calculating PDFs in this paper.  Roughly, the recurring
idea in all the variants of the perturbative matching approaches
is the factorization of certain equal-time boosted hadron matrix
elements with a spatial operator separation $z$ (or the Fourier
transform thereof into $x$-space) into a matching kernel and the
structure functions, such as the PDF as a specific case.  The
usefulness of this factorization is that the matching kernel can
be perturbatively evaluated and hence universal for all hadrons.
We refer the reader to the recent
reviews~\cite{Constantinou:2020pek,Cichy:2018mum,Ji:2020ect,Radyushkin:2019mye}
on this broad topic for further details on the methodology and for
the recent works in this direction.

Without any loss of generality, in this paper, we will specifically
consider the case of equal-time multiplicatively
renormalizable~\cite{Ishikawa:2017faj,Ji:2017oey} correlator
constructed from the quasi-PDF~\cite{Ji:2013dva} matrix element
\beq
2 E(P_z) {\cal M}(\om, z^2) = \langle P_z | \bar\psi_z \gamma_t W_{z,0} \psi_0 |P_z \rangle,
\eeq{itdef}
of quark and antiquark separated by a spatial distance $z$, also
known as pseudo-Ioffe-time distribution~\cite{Radyushkin:2017cyf}
(abbreviated as pseudo-ITD or just as ITD), and evaluated in an
on-shell hadron with $P = (E, P_z)$ moving with a spatial momentum
$P_z$. The variable $\om=z P_z$ is also known as the
Ioffe-time~\cite{Ioffe:1969kf,Braun:1994jq}.  However, the rest of
the paper can be equally carried over to any of the operators
(e.g.,~\cite{Sufian:2019bol,Sufian:2020vzb}) under the umbrella of
the good lattice cross-sections as well.  The pertinent detail for
this paper is that the the underlying factorization of ${\cal M}(\om,
z^2)$, can be equivalently rewritten~\cite{Braun:2007wv,Izubuchi:2018srq}
as a leading twist operator product expansion (OPE),
\beq
{\cal M}(\om,z^2) = \sum_{n=0} C_n(z^2\mu^2) \frac{(-i \omega)^n}{n!} \langle x^n\rangle(\mu)+{\cal O}(\Lambda_{\rm QCD}^2 z^2),
\eeq{ope}
valid up to higher-twist corrections, usually denoted as ${\cal
O}(\Lambda_{\rm QCD}^2 z^2)$ terms.  Occupying the central role in
the above expression are the Wilson coefficients $C_n(z^2\mu^2)$
that match the renormalized Euclidean matrix element in
some scheme to the $\msbar$ PDF via the moments, and can be
perturbatively computed at the hard scale set by $1/z$. In a typical
lattice computation, the analytic form of $C_n$, 
\beq
C_n(\mu^2 z^2) = \tilde{c}_0(n) + \tilde{c}_1(n) \ln(\mu^2 z^2) + \tilde{c}_2(n)\ln^2(\mu^2 z^2)+\ldots,
\eeq{cn}
in terms of the
coefficients $\tilde{c}$ of the powers of the singular logarithms $\ln(\mu^2 z^2)$,
up to certain order in $\alpha_s$ is provided as an input,
and in return, the PDF $f(x,\mu)$ and the Mellin moments, $\langle
x^n\rangle = \int_0^1 x^n f(x) dx$, are computed as outputs~\footnote{
For the flavor non-singlet PDF case that we will specifically consider in this paper, the
notation for the Mellin moments as given in the community white paper~\cite{Lin:2017snn} are
$\langle x^n \rangle_{u^+-d^+}$ for the odd values of $n$, and 
$\langle x^n\rangle_{u^--d^-}$ for the even values of $n$ that corresponds to
the valence PDF moments. We will simply refer to these non-singlet moments, specifically the 
valence moments, as $\langle x^n\rangle$ throughout this paper.}.

In the rest of the paper, we will consider the non-singlet,
valence parton distribution functions. Unlike the singlet
distributions, for the valence case, $x f(x)$ vanishes in the
small-$x$ limit, and hence, enable us to neglect the associated
systematic errors in the global fit determinations and in their first
few Mellin moments. This lets us focus only on sources of uncertainty in
the lattice computations of PDF. 
In the valence sector, the perturbative matching approach has been successfully
applied to the lattice computations of PDF
of the pion~\cite{Sufian:2019bol,Sufian:2020vzb,Izubuchi:2019lyk,Gao:2020ito,Lin:2020ssv}
and the proton~\cite{Joo:2020spy,Joo:2019jct,Orginos:2017kos,Bhat:2020ktg,Fan:2020nzz,Alexandrou:2020qtt,Alexandrou:2018pbm,Lin:2020fsj}. In these calculations, the resulting reconstructed PDFs capture
more or less the features of the phenomenologically determined PDFs
(which we henceforth simply refer to as ``pheno-PDFs"). By avoiding
the issues of inversion problem to obtain $x$-dependent PDF, a
direct comparison of the lattice ITD data itself with the corresponding
expectation from the global analysis fits, obtained via perturbative
matching, have only been qualitatively {\sl close enough}  (for
example, Ref.~\cite{Bhat:2020ktg} being one such comparison with
NNPDF result~\cite{Ball:2017nwa} for the proton at physical point,
and Ref.~\cite{Gao:2020ito} for comparison of JAM18
result~\cite{Barry:2018ort} with 300 MeV pion). In particular, the
deviation between the lattice ITD and the pheno-ITD increases with
$\om$ or $z$ as discussed in~\cite{Sufian:2020wcv}. These presumably
point to the presence of yet uncontrolled systematic effect(s).
Some of the reasons for this could be:

\begin{enumerate}

\item The effect of heavier than physical pion masses used in most
of the calculations, but as we mentioned
above, a recent computation~\cite{Bhat:2020ktg} of the isovector
nucleon PDF at the physical point still shows visible discrepancy
with the ITD constructed from NNPDF global analysis~\cite{Ball:2017nwa}.
The weak pion mass dependences seen in~\cite{Joo:2020spy, Sufian:2020vzb}
in the range,  $m_\pi \in [172,415]$ MeV, suggests this might not be the
dominant contribution to the observed discrepancies even for
computations at moderately heavier pion masses.

\item  There could be lattice spacing corrections to pseudo-ITDs.
For the case of pseudo-ITD renormalized in the ratio scheme, the
lattice spacing effects were seen~\cite{Gao:2020ito} to mostly
affect regions of ${\cal M}(\om,z^2)$ for $z={\cal O}(a)$.  Such
lattice spacing effect at small $z$ was also seen for the nucleon
recently~\cite{Karpie:2021pap}. Therefore, at least for renormalization
methods like the ratio, the short-distance lattice spacing effects
are less important and can be ignored provided one skips first few
values of $z$. Hence, we do not consider this in this paper.

\item A more troublesome reason could be the presence of yet
undetected large higher-twist contributions in the range of
separations, $z \in [0.04, 1]$ fm that are accessible in the
computations today, but incorrectly taken as part of leading-twist
contribution. However, a recent investigation~\cite{Gao:2020ito}
of the higher-twist effect that dominate in the $P_z=0$ RI-MOM 
matrix element for the pion suggests that the higher-twist effect
could be a minor, $\approx 40$ MeV effect that is atypically small
compared to the QCD scale, $\Lambda_{\rm QCD}\approx 300$ MeV.  By
using ratio method (with ratio taken with respect to zero
momentum as well as with non-zero momentum bare matrix element)
for renormalization instead of the RI-MOM
scheme~\cite{Chen:2017mzz}, it is reasonable to expect these leading
$z^2$ higher-twist contributions to get approximately canceled
leaving behind only a residual higher-twist contamination of the
form $z^2 \om^2$~\cite{Karpie:2018zaz}.

\end{enumerate}

In this work, we take an optimistic stand on this matter and ask
what if the remaining uncertainties in the leading-twist approach
are predominantly unaccounted higher-loop corrections, and how to 
take care of our limited knowledge of the perturbative convergence of 
the Wilson coefficients in a probabilistic manner? This is the
motivation for this paper. 

There have been previous works to find ways to take into account
the residual perturbative corrections and consider the question of
whether the lattice results have perturbatively converged i.e.,
whether $C_n(z^2)$ is perturbatively convergent at all potential
values of $z$ that are used in a typical lattice analysis.  There
is an ongoing program to find higher loop corrections to $C_n$,
with the state-of-the-art result at next-to-next-to-leading order (NNLO) published
in~\cite{Chen:2020arf,Chen:2020iqi,Li:2020xml}.  In a recent
paper~\cite{Gao:2021hxl}, the residual effect of higher-loops
partially seen via resumming next-to-leading logarithms and threshold
resummation was investigated in detail, wherein visible effects of
resummation were seen in typical values of sub-fermi distances $z$.
For the sake of argument, in addition to the resummation of
lower-order logarithms, {\sl a priori} it cannot be ruled out that
the genuine higher-loop corrections and logarithmic terms that arise
via 1PI diagrams could result in large numerical coefficients that
multiply $\alpha_s^m$ factors, making the convergence of perturbation
theory slower for the leading-twist approach.  A method to take
care of the renormalon ambiguity in the perturbation series was
considered recently in~\cite{Ji:2020brr}.

In this paper, we assume two things: first, that the lattice data
for the matrix element has a leading-twist description. Second, we
assume that the Wilson coefficients that enter the leading-twist
OPE are hadron independent.  Given these two assumptions, the idea
we pursue in this paper is to replace a perturbatively determined
$C_n$ with a probabilistically likely $C_n$ that is determined
directly from the lattice matrix element data, and given a trustworthy
knowledge of valence PDF of a hadron from global analysis of
experimental input.  The lattice determined $C_n$ can then be used
in computing the PDFs of other hadrons, as well as to obtain the
zero-skewness GPD. By turning the argument around, as this Bayesian
approach is falsifiable, it would also be a good way to rule out
the hypothesis of higher-loop corrections being important in matching
the lattice calculations (which is the premise of this paper), or
whether there are substantial higher-twist corrections affecting
the lattice results.  Therefore, at the cost of adding the systematics from the lattice computations for
two different hadrons, the method presented in this paper
will help disentangle the perturbative QCD systematics from systematics of nonperturbative 
lattice parton computations.
One should note that our method is a novel
way to make use of global analysis data in lattice computation,
unlike the previous
ways~\cite{Bringewatt:2020ixn,DelDebbio:2020rgv,Cichy:2019ebf,Lin:2020rut},
where one tries to include the lattice data as one of the input to
global analysis.  In the rest of the paper, we explain the method
and apply it to lattice mock-data, and some real published proton
and pion lattice data.

\section{The method}

In the traditional Bayesian reconstruction of PDFs $f(x)$ via fits
(of parameters that characterize a certain PDF ansatz) to the
lattice data for the matrix element ${\cal M}_H(\om,z^2)$ for a hadron $H$, 
one asks for the conditional probability distribution
\beq
P\bigg{(}f(x)\bigg{|}\mathrm{data\ for\ } {\cal M}_H\bigg{)} \propto P\bigg{(}\mathrm{data\ for\ } {\cal M}_H\bigg{|}f(x)\bigg{)} P\bigg{(}f(x)\bigg{)}.
\eeq{bayes1}
One treats $P(\mathrm{data}|f)\sim e^{-\chi^2/2}$ via the
$\chi^2$-minimization analysis to find the best fit value (i.e., the 
maximum a posteriori estimator) of $f(x)$
that maximizes the left-hand side
probability, and its confidence interval.  The assumption of a certain parametrization of the
PDF is implicit in the prior $P(f)$.  Methods based on the Bayesian
reconstruction of PDFs so as to avoid the modeling bias in $f$ have
been tried before~\cite{Karpie:2019eiq}.
Instead of applying the Bayesian methods to the reconstruction of
PDFs, in this work, we apply the same ideas to the determination of
Wilson coefficients. We propose the following two-step
scheme assuming one has lattice results for more than one hadron
--- results for ${\cal M}_H(\om,z^2)$ for a hadron $H$ (say, for
the proton), and the lattice data for ${\cal M}_{H'}(\om,z^2)$
for at least one additional hadron $H'$ (say, for the pion). 

\subsection{Step-1: Fit the Wilson coefficients given the lattice
data for a hadron and prior knowledge of its pheno-PDF}

In the first step, we ask for the conditional probability distribution
$Q$ for the set of $z$- and $n$-dependent Wilson-coefficients,
$\{C_n\} = \left\{ C_n(z) \text{\ for\ } z\in[a,z_{\rm max}]  \text{\
and\ } n\in [0,N_{\rm max}]\right\}$, themselves, given the lattice data
for the matrix element ${\cal M}_H(\om,z^2)$ for a hadron $H$, as well
as the prior knowledge of the PDF $f_H(x)$ of the same hadron $H$ from
phenomenological determinations,
\beqa
Q(\{C_n\})&\equiv& P\bigg{(}\{C_n\}\bigg{|}\mathrm{data\ for\ } {\cal M}_H, f_H \bigg{)}, \cr &\propto& P\bigg{(}\mathrm{data\ for\ } {\cal M}_H \bigg{|} \{ C_n\}; f_H\bigg{)} P\bigg{(}\{C_n\}\bigg{)}.\cr&&\quad
\eeqa{newbayes1}
As is standard, the distribution $P(\mathrm{data\ for\ } {\cal M}_H| \{ C_n\}; f_H)$ models the 
Gaussian fluctuation of the data around the theoretical model ${\cal M}^{\rm twist-2}(\om,z; \{C_n\}, f_H)$.
That is, we can conveniently define, $Q(\{C_n\}) ={\cal K}\exp(-\chi^2/2)$, with a normalizing constant ${\cal K}$
and 
\beqa
&&\chi^2\left(\{C_n\}\right) = \sum_{z, \om}\frac{\left( {\cal M}_H(\om,z) - {\cal M}^{\rm twist-2}(\om,z; \{C_n\}, f_H)\right)^2}{\sigma_H^2}\quad\cr &&\qquad\quad\quad\quad -2 \ln\left(P(\{C_n\})\right),
\eeqa{chi2}
where, $\sigma_H$ are the statistical errors associated with the lattice data for ${\cal M}_H(\om,z)$. In our case, 
the theoretical model of the data is the twist-2 OPE that is dependent on the Wilson-coefficients as 
well the PDF (via its moments),
\beq
{\cal M}^{\rm twist-2}(\om,z; \{C_n\}, f_H) = \sum_{n=0}^{N_{\rm max}} C_n(z) \langle x^n\rangle_H \frac{(-i\om)^n}{n!},
\eeq{twistope2}
with the Mellin moments, $\langle x^n\rangle_H = \int_0^1 x^n f_H(x) dx$ as
fixed numbers~\footnote{In this paper, we will treat the valence pheno-PDF as if they have 
no statistical errors. It can be incorporated in a full-fledged lattice analysis by using the available PDF data sets.}.  
Crucial to the Bayesian framework is the prior distribution $P(\{C_n\})$. 
For example, one could impose
a non-informative prior on $C_n$, in which case $P(\{C_n\})$ is a
constant for any choice of $\{C_n\}$, and therefore irrelevant.  
An alternative is to
impose the degree of our confidence in the perturbative estimate (determined 
up to certain order), $C_n^{\rm pert}(z)$, through a Gaussian prior distribution,
\beq
-\ln\left(P(\{C_n\})\right) = \sum_{z,n} \left(\frac{\left(C_n(z) - C_n^{\rm pert}(z)\right)^2}{2\sigma_{\rm prior}^2}\right),
\eeq{prior}
where the sum runs over all $z$-values
and all the values of order $n$ in the twist-2 OPE that are included
in the analysis.  For example, one could use a broad prior width
$\sigma_{\rm prior}$ to input no bias, or another
choice could be two or three times the variations coming from
changing the scale from $\mu/2$ to $2\mu$ used to determine $\alpha_s$
that enters $C_n^{\rm pert}(z)$. 

This first step to characterize $Q(\{C_n\})$ can be implemented in different 
ways depending on the computational effort that one is willing to invest. In 
perhaps the most accurate implementation, one can simulate the multi-dimensional 
integral, ${\cal K} \int e^{-\frac{1}{2}\chi^2(\{C_n\})} {\cal D} C_n$,
by using the Markov-Chain Monte Carlo (MCMC) numerical simulation. Thereby, one can 
estimate averages with respect to the distribution $Q$, 
\beq
\left\langle F\right\rangle_{C_n} \equiv \int F(\{C_n\}) Q(\{C_n\}) {\cal D}C_n, 
\eeq{qmean}
of 
functions $F(\{C_n\})$ that depend on the Wilson coefficients. The
Expected A Posteriori (EAP) values of $C_n$  obtained by using 
$F(\{ C_n\}) = \{C_n\}$ is one such example that can computed using MCMC. A cheaper 
alternative to using MCMC is to first summarize the information in $Q(\{C_n\})$ using the
Maximum A Posteriori (MAP) values of $\{C_n\}$ that maximizes $Q(\{C_n\})$. That is 
\beq
\{C_n^{\rm MAP}\}\equiv \argmax{\{C_n\}} Q(\{C_n\}) = \argmin{\{C_n\}} \chi^2(\{C_n\}),
\eeq{mapvals}
which can be obtained by the standard $\chi^2$-fits to the data by minimizing 
\eqn{chi2}. The confidence intervals $C_n^{\rm MAP}\pm \sigma_{C_n}$ can be 
obtained by bootstrap. With the MAP values, a cheaper possibility to sample the distribution $Q$ 
is to approximate it
parametrically by the multinormal distribution ${\cal N}\left(\{C_n^{\rm MAP}, \sigma_{\rm C_n}\}\right)$.
Instead of using a multinormal distribution, 
one could use the non-parametric bootstrap histogram of $\{C_n^{\rm MAP}\}$
obtained by performing the $\chi^2$ minimization in each of the bootstrap samples of the data.
For demonstration purposes, we will use this simpler procedure in the latter half of paper.
In these approximations to the true $Q(\{C_n\})$, it is important to resample
the central values $C_n^{\rm pert}$ entering the prior distribution from 
$P(\{C_n\})$ in each bootstrap iteration, so as to ensure that in the absence of any constraint from the data, 
one reproduces the prior distribution $P(\{C_n\})$ rather converging to the same value in every iteration.
For the sake of completion on the possible implementations of Step-1, as a 
crude approximation, one could simply ignore the uncertainty in $C_n^{\rm MAP}$ and use
$Q(\{C_n\})\approx \delta(\{C_n\}-\{C_n^{\rm MAP}\})$ which will greatly simply the analysis steps above and 
those that follow. However, we do not follow this approximation here.

\subsection{Step-2 : Determine the PDF of other hadrons with the computed Wilson coefficients}

In the second step, we use the estimated $C_n$ from
the first step to fit the ${\cal M}_{H'}(\om,z^2)$-data for
another hadron $H'$.  That
is,  we ask for the conditional probability on the parameters $\alpha$
describing the PDF $f_{H'}(x;\alpha)$ of $H'$ given the data and marginalized over the random choices
of $\{C_n\}$ picked from the distribution $Q(\{C_n\})$,
\beqa
&& P\bigg{(} f_{H'}\bigg{|}\mathrm{data\ for\ } H'\bigg{)}  \propto \cr && P(\alpha) \int P\bigg{(}\mathrm{data\ for\ } H'\bigg{|} f_{H'}, \{C_n\}\bigg{)}  Q( \{ C_n\}) {\cal D} C_n .\cr&&\quad
\eeqa{newbayes2}
The distribution, $P(\alpha)$, on the right-hand side is the prior on the parameters $\alpha$
that enter $f_{H'}$, given a fixed underlying model for the PDF.
For example,
$f_{H'}(x;\alpha)$ could be the model $x^a(1-x)^b$, with $\alpha=\{a,b\}$.
In the rest of the paper, we will consider a non-informative prior on $\alpha$, and 
hence, drop the prior term $P(\alpha)$ for the sake of simplicity.
The distribution, $P\left(\mathrm{data\ for\ } H'| f_{H'}, \{C_n\}\right)\equiv {\cal K}' e^{-\chi'^2(\alpha,\{C_n\})/2}$ with the normalizing constant ${\cal K}'$,
captures the Gaussian fluctuation of data for ${\cal M}_{H'}$ around the twist-2 OPE theoretical model via
\beqa
&&\chi'^2\left(\alpha,\{C_n\}\right) =\cr&&\quad \sum_{z, \om}\frac{\left( {\cal M}_{H'}(\om,z) - {\cal M}^{\rm twist-2}(\om,z; \{C_n\},\alpha)\right)^2}{\sigma_{H'}^2}.
\eeqa{chi2forfh}
To be concrete, one is interested in a measure of the most probable
value of $f_{H'}$ and its confidence intervals. Similar to the
discussion in Step-1, one such quantity that can be measured in an
MCMC simulation is the EAP value of the PDF,
\beq
f_{H'}^{\rm EAP}(x) ={\cal K}' \int f_{H'}(x;\alpha) e^{-\frac{1}{2}\chi'^2(\alpha,\{C_n\})}  Q( \{ C_n\}) {\cal D}\alpha {\cal D} C_n.
\eeq{pdfeap}
To evaluate the above integral, one will sample the values of
$\{C_n\}$ from the MCMC simulation as performed in Step-1 in order
to capture the measure $Q(\{C_n\}){\cal D}C_n$, and perform a second
MCMC simulation over the parameters $\alpha$. This would be
computationally rather complex. Instead, as a simpler statistic for
the probable value of $f_{H'}(x;\alpha)$, one can find the Maximum
Likelihood Estimate (MLE), $\left[f_{H'}\right]^{\rm MLE}({C_n})$,
that minimizes $\chi'^2(\alpha,\{C_n\})$ for each instance of
$\{C_n\}$ drawn from $Q(\{C_n\})$, and then perform an average over
$C_n$. That is, one can use
\beq
f^{\rm avg}_{H'} \equiv \left\langle \left[f_{H'}\right]^{\rm MLE}\right\rangle_{C_n},
\eeq{newmean}
as an estimator of the PDF and its confidence intervals.
To perform the average $\langle \ldots\rangle_{C_n}$, one needs to draw instances of $C_n$ sampled in 
Step-1. As mentioned previously, we will use the bootstrap sample of $\{C^{\rm MAP}_n\}$ from Step-1 as an 
approximation of $Q(\{C_n\})$ to perform the above step. To avoid extra symbols, we will simply 
use $f_{H'}(x)$ to denote $f^{\rm avg}_{H'}(x)$ and its error-bands in the rest of the paper.

The two steps could 
be done in such a way that collaboration-A does a dedicated
computation of an ensemble of $\{C_n\}$ distributed according to 
$Q(\{C_n\})$ in its own set of gauge ensembles etc., using some hadron $H$,
and collaboration-B takes this $Q(\{C_n\})$ as an input for
its dedicated computation of $f_{H'}(x)$ for hadron $H'$. In this
case, there is no correlation between a choice of $C_n'$ and the
re-sampled bootstrap configurations.  In another way, a single lattice
collaboration computes both pion and proton, preferably at the physical point and using the same 
set of configurations. In this case, one could reduce the statistical errors by sampling $Q(\{C_n\})$ 
so as to maintain correlations between the pion and proton data.

\subsection{Remarks on the method}

First, we have a choice as to which hadron we should consider first
to determine the Wilson coefficients. If one has very accurate data,
this question simply boils down to how well determined the pheno-PDFs
of a given hadron is.  For the unpolarized sector, using the proton
valence PDF, which has been determined at NLO and NNLO level
accuracy~\cite{Ball:2017nwa}  from the experimental data, as the known
input would be the ideal choice. One can then input the set of $C_n$
from the proton analysis into the analysis of pion PDFs on the
lattice.  The downside to this order of steps is that the moments
$\langle x^n\rangle$ of proton PDF approach zero faster as $n$ is
increased, compared to the pion~\footnote{The reason being related
to the large-$x$ behavior $f\sim (1-x)^\beta$ of the PDFs, and one
expects asymptotically $\langle x^n\rangle\sim n^{-\beta}$ with
$\beta$ for proton being larger than pion.}. Consequently, while
one can determine $C_n(z)$ for to a higher order in $n$ (we expect
about up to first three even $n$, as we will discuss later in the
paper) using pion PDF as input, it would be difficult to do so using
the proton. Nevertheless, this difficulty is purely statistics related and can be 
solved in a high-statistic study of the proton ITD in the future.
On the other hand, the difficulty in using valence pion PDF as
the input PDF is that unlike the proton PDF, the robustness of valence
pion PDF from different analyses of experimental
data~\cite{Badier:1983mj,Betev:1985pf,Conway:1989fs,Owens:1984zj,Sutton:1991ay,Gluck:1991ey,
Gluck:1999xe, Wijesooriya:2005ir,Aicher:2010cb,Barry:2018ort,Cao:2021aci}
are still being scrutinized.  In this paper, to simply demonstrate
the idea, we will use the pion PDF as the starting point, but use
different versions of pheno-PDFs of pion as inputs in the lattice
analysis, and then determine the $C_n$ for each such choice. 
In turn, this procedure will lead to different estimates of proton PDFs on the
lattice.  In this way, in addition to demonstrating the method, we
would also be studying the impact of a good phenomenological
determination of pion-PDF on the lattice determination of proton
PDFs.

Secondly, let us elaborate on the method being a Bayesian extension
of the traditional analysis of lattice PDFs. As our confidence in
the perturbative Wilson coefficients increases, due to say its
calculation up to a certain higher-order $m$, then by the Bayesian
framework, one will choose the prior distribution that peaks more
and more about the $m$-loop result for $\{C_n\}$. That is, one will
choose $\sigma_{\rm prior}\to 0$ as the confidence is subsequently
increased. In that case, the posterior distribution approaches a
delta-function about the $m$-loop result, and one recovers back the
usual way of analysis using a given perturbative $C_n$.  This is
the essence of the idea in this paper to treat our ignorance of
$C_n$ in a probabilistic manner.

Thirdly, it is easy to see that the fitted coefficients will
automatically satisfy the Altarelli-Parisi evolution to the same
accuracy to which the input pheno-PDF moments satisfies the evolution,
since the matrix element ${\cal M}_H$, that is renormalized in a lattice scheme 
such as the ratio, is independent of the
factorization scale $\mu$. To expand on, the actual quantities that
we obtain by fits are the $z$-dependent coefficients of the $(-i\om)^n/n!$-terms 
in the twist-2 OPE in \eqn{twistope2}. Let us call those
coefficients $D_n(z)$.  Since ${\cal M}_H$ has no knowledge of
$\mu$, the coefficients $D_n(z)$ will be constants independent of
$\mu$. By writing $D_n(z) = C_n(\mu,z) \langle x^n \rangle(\mu)$,
it is clear that the $\mu$-dependence of fitted $C_n(\mu,z)$ will
be the inverse $\mu$-dependence of $\langle x^n \rangle(\mu)$, which is dictated by
the Altarelli-Parisi equation that the global fit PDF satisfies.

Fourthly, since we are using only the valence Mellin moments in 
the twist-2 OPE to determine $C_n$, it is not necessary to use the 
pheno-PDF values, and instead, one could use lattice determinations of Mellin 
moments obtained using the local twist-2 operators (e.g.,~\cite{Mondal:2020cmt,Alexandrou:2021mmi}
for recent studies). Using the local twist-2 operator approach, 
both even and odd Mellin moments can be computed. However, currently, only $\langle x^2\rangle$ even valence moment
from such studies are usable as input to determine $C_n$ in the method we proposed (and 
only even $n$ enter the valence ITDs), and hence not a practical option currently.

\section{Generalization to zero-skewness GPD}

Let us now consider the equal-time matrix element corresponding to
the generalized parton distribution function (GPD) $f(x, t)$ at
zero-skewness~\cite{Liu:2019urm,Radyushkin:2019owq},
\beq
2 E {\cal M}(\om, z^2, t=-q^2) = \langle \vec{P}+\vec{q}/2 | \bar\psi_z \gamma_t W_{z,0} \psi_0 | \vec{P}-\vec{q}/2  \rangle,
\eeq{gpd}
with spatial momentum $\vec{P} = (0, 0 , P_z)$ and transverse
momentum transfer $\vec{q} = (q_x, q_y, 0)$, with $t=\vec{q}^2$.
For this kinematics at zero skewness, $\xi = q_z/2P_z = 0$, the
corresponding leading twist OPE is given as~\cite{Liu:2019urm,Radyushkin:2019owq} ,
\beq
{\cal M}(\om,z^2, t) = \sum_{n=0} C_n(z^2\mu^2) \frac{(-i \om)^n}{n!} \langle x^n\rangle(t, \mu),
\eeq{opegpd}
with the same set of Wilson coefficients $\{C_n\}$ as in the case of
forward PDF~\cite{Liu:2019urm}\footnote{We thank Yong Zhao for
clarifying that this property is expected to all orders in perturbation
theory.}, but the GPD moments (i.e., the generalized form factors), $\langle x^n\rangle(t, \mu)$ being
dependent on the momentum transfer $t$. Given this property of the
$\xi=0$ GPD, it is straight forward to generalize the analysis
scheme presented in the last section:
\begin{enumerate}
\item Determine $Q(\{C_n\})$ using the lattice forward matrix element
data ${\cal M}_H(\om, z^2, t=0)$ of a hadron $H$ using the pheno-PDF
$f_H(x)$ as a given. This step is the same as the step-1 described
previously.

\item Using randomly sampled $\{C_n\}$ from the distribution $Q(\{C_n\})$,  perform the analysis of zero
skewness GPD of the same hadron $H$ at $t>0$ via $P\bigg{(}
        f_H(x,t)\bigg{|}\mathrm{data\ for\ } {\cal M}_H(t)\bigg{)}$ that is marginalized over 
        $\{C_n\}$.

\end{enumerate}
We expect this to be a less-noisy implementation as the GPD data
of a hadron $H$ is expected to be correlated with the $t=0$
matrix element of the same hadron $H$ in a given ensemble.  It
would be interesting to see the data presented for valence pion
GPD~\cite{Chen:2019lcm} and unpolarized proton
GPD~\cite{Alexandrou:2020zbe} analyzed in the manner suggested here.

In addition to the generalization to zero-skewness GPD, few other
cases are also possible~\footnote{We thank the anonymous referee
for pointing us to the additional possibilities.}.  The chiral-symmetry present
in the massless quark limit ensures that the Wilson coefficients
are the same for bilocal quark bilinear operators with $\gamma_z$
and $\gamma_5\gamma_z$ Dirac spinor structures~\cite{LatticeParton:2018gjr,Fan:2020nzz}. That is, the matching
factors are the same for the unpolarized PDF computed with $\gamma_z$
bilocal quark bilinear and  the helicity PDF computed with
$\gamma_5\gamma_z$ bilocal bilinear. Thus, the method presented
here can be applied to the proton unpolarized PDF determined using
$\gamma_z$ bilocal bilinear to determine the corresponding $\{C_n\}$,
and then use them in the lattice computation of the proton helicity PDF. The
drawback is that the $\gamma_z$ bilocal bilinear suffers from
mixing with the scalar bilinear~\cite{Constantinou:2017sej,Chen:2017mzz}, adding to the systematics.
In addition, the Wilson-coefficients for $\gamma_z$ bilocal bilinear
also enters the leading-twist OPE that describes the Distribution
Amplitude (DA) of the pion~\cite{Braun:2007wv,Liu:2018tox,Radyushkin:2019owq}.

\section{Demonstration of the method using mock-data}

\bef
\centering
\includegraphics[scale=0.8]{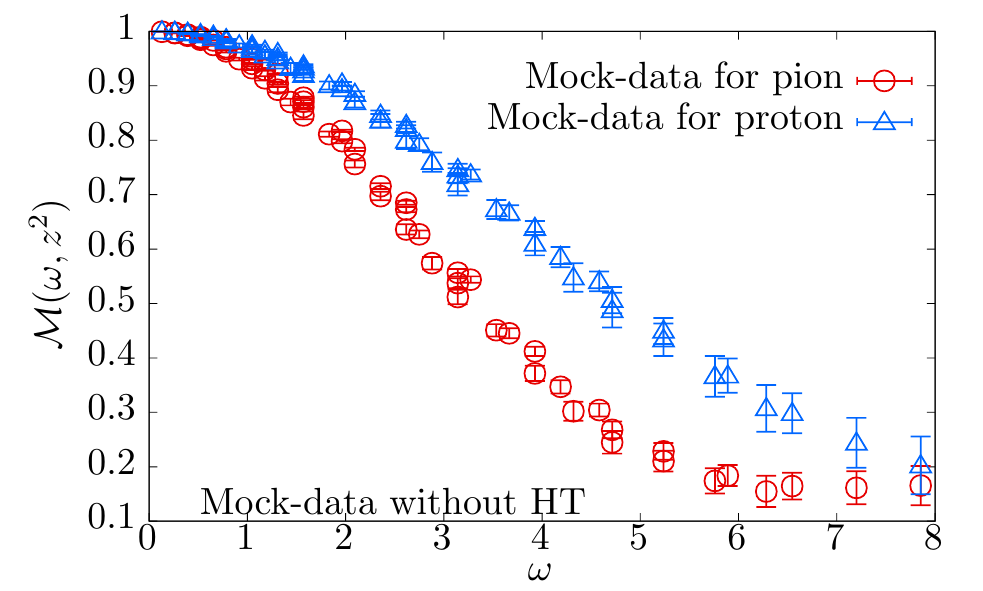}
\caption{A mock-data for the real part of the pseudo-ITD for pion (red) and proton (blue) shown as a function of 
$\om=zP_z$. The data from $z\in [a, 12a]$ have been put together in the plot (and not distinguished 
by their $z$ values to avoid clutter), leading to visible corrections to the scaling with $\om$ and will be 
captured by the Wilson coefficients $C_n(z)$. For parameters used to generate this data, refer 
text.}
\eef{mockdemo1}

In this section, we apply the idea to a set of mock-data that we generated
for the pion and proton pseudo-ITD ${\cal M}(\om, z^2)$, which are renormalized in the ratio scheme
and set in a lattice with a spacing
$a=0.06$ fm with an extent $L_z=48$ in the $z$-direction.  Also,
we will only be looking at the real part of the ITD, ${\rm Re} {\cal
M}$, which corresponds to the nonsinglet valence contribution for
both the proton and the pion.  The point of using mock-data is that
we exactly know what corrections to NLO matching we put in, and ask
how this new framework will handle these corrections.  We generated
the mock-data for pseudo-ITD of $H$ (proton $p$ and pion $\pi$) by
drawing sample means $\overline{{\cal M}}_H(z P_z, z)$, for $z\in
[a, 12a]$ and $P_z aL_z/(2\pi) \in [1, 5]$. The underlying distribution
for the sample means $\overline{{\cal M}}_H$ (which one can think
of as column vector of size 60 by linearizing $\om$ and $z$) was
the multinormal distributions, ${\cal N}\left(\overline{{\cal
M}}^0_H, \Sigma\right)$.  The true central value $\overline{{\cal
M}}^0_H$ of the multinormal distribution was engineered by hand so
as to have corrections to NLO values of $C_n$, and certain amount
of higher-twist correction, and written as
\beqa
\overline{{\cal M}}^0_H(\om,z^2) &=& \sum_{n=0} C^{\rm mock}_n(z^2\mu^2) \frac{(-i \om)^n}{n!} \langle x^n\rangle_H  \cr && +\quad\left(e^{-\Lambda^2 z^2 \om^2} -1 \right),
\eeqa{opemock}
where the moments $\langle x^n\rangle_H$ were inferred from the
pheno-PDF values of $H$; namely, from the unpolarized NNPDF~\cite{Ball:2017nwa}
data at $\mu=3.2$ GeV for the proton, and the JAM20~\cite{Cao:2021aci,Barry:2018ort} data at
$\mu=3.2$ GeV for the pion respectively. Since we are looking at the real part of the ITD, only the 
even $n$-values enter the above equation. We use a set of mock
Wilson coefficients $C_n^{\rm mock}$ given by the coefficients of
the $\ln^j(\mu^2 z^2)$ terms,
\beq
\tilde{c}_n(j) = \tilde{c}^{\rm NLO}_n(j)  + \Delta \tilde{c}_n(j),\quad j=0,1,2,
\eeq{mockcn}
used in \eqn{cn}. Above, $\tilde{c}^{\rm NLO}_j$ are the NLO values
that can be read off from~\cite{Izubuchi:2018srq,Karpie:2018zaz}, and $\Delta \tilde{c}_j$ are
{\sl toy} correction terms that we put in by hand to simulate the
higher-loop corrections. Since we are working with ratio matrix
elements, we simply take $C_0^{\rm mock}(\mu^2 z^2)=1$, as it is
true to all orders. We added the residual higher-twist correction
via the term in the second line of \eqn{opemock}; it is motivated
by its expansion as $\Lambda^2 z^2 \om^2 +\Lambda^4 z^4 \om^4/2+\ldots$,
which involves all $z^k$ type terms, and we have chosen it as a
function of $z^2 \om^2$ as any leading $z^2 \om^0$ type higher-twist
effect would most-likely have been canceled in the ratio scheme.
As for the covariance matrix $\Sigma$ in the multinormal distribution,
we made an ad hoc choice $\Sigma_{[z,\om], [z',\om']} = \sigma_{z,\om}
\sigma_{z',\om'}/(1 + k |\om-\om'|^\delta)$, with the diagonal term
being simply the $\sigma_{z,\om}^2$ error term, and the $|\om-\om'|^\delta$
term controlling how the correlation in the mock-data falls off as one
moves along the off-diagonal elements; we chose $\delta=0.5$ and
$k=0.5$ for maintaining some correlation in the mock-data. As for the diagonal error
term, we chose a model $\sigma_{z,\om} = r |\om z|$ motivated by
the fact that for a given value of $P_z$, the data gets
noisier as $z$ increases, and that at a fixed $z$, the data at a
higher $P_z$ is noisier. We used $r=0.0005$ for the pion and 0.0008 for the
proton. The only justification for choosing all these models to
generate mock-data is simply that it leads to reasonable looking
data that is similar to what is seen in the published literature.

\befs
\centering
\includegraphics[scale=0.72]{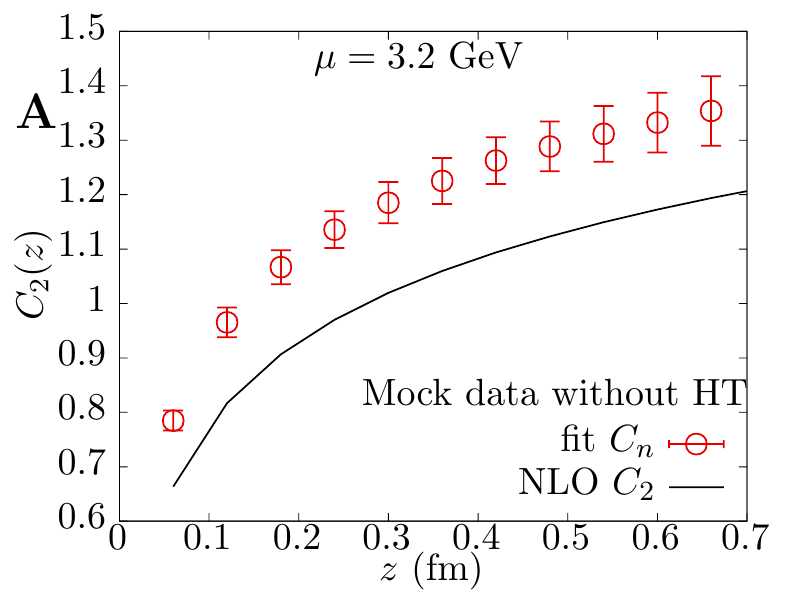}
\includegraphics[scale=0.72]{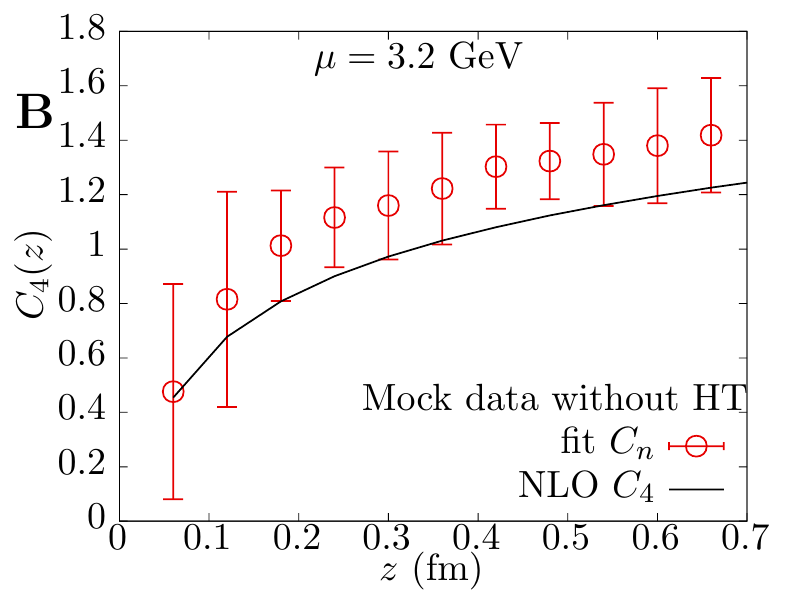}
\includegraphics[scale=0.72]{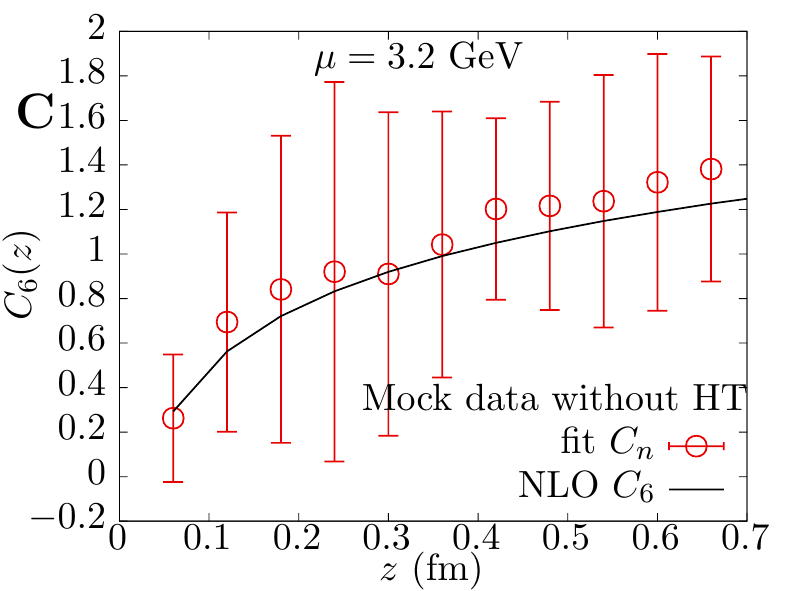}

\includegraphics[scale=0.72]{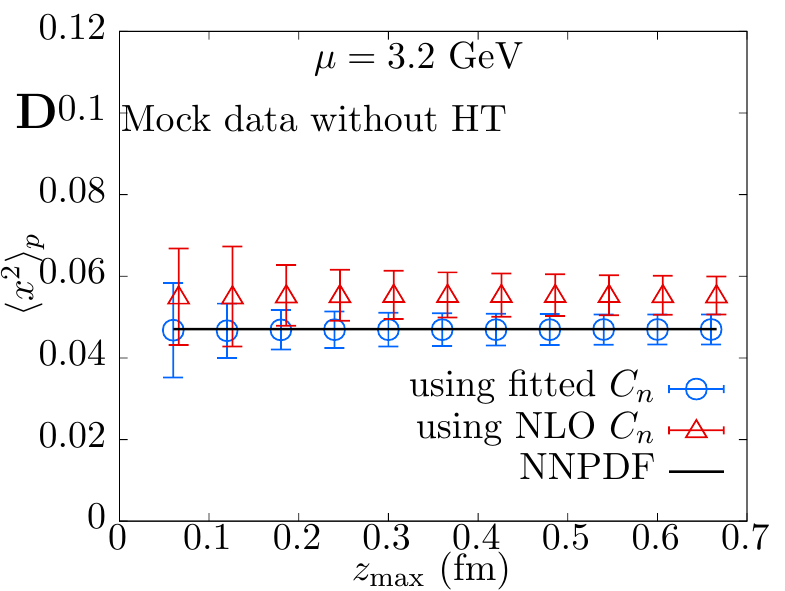}
\includegraphics[scale=0.72]{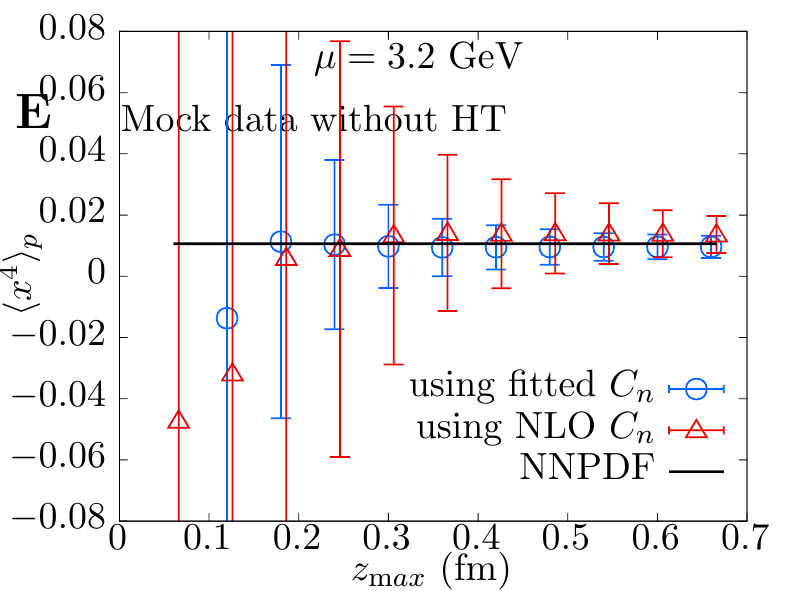}
\includegraphics[scale=0.72]{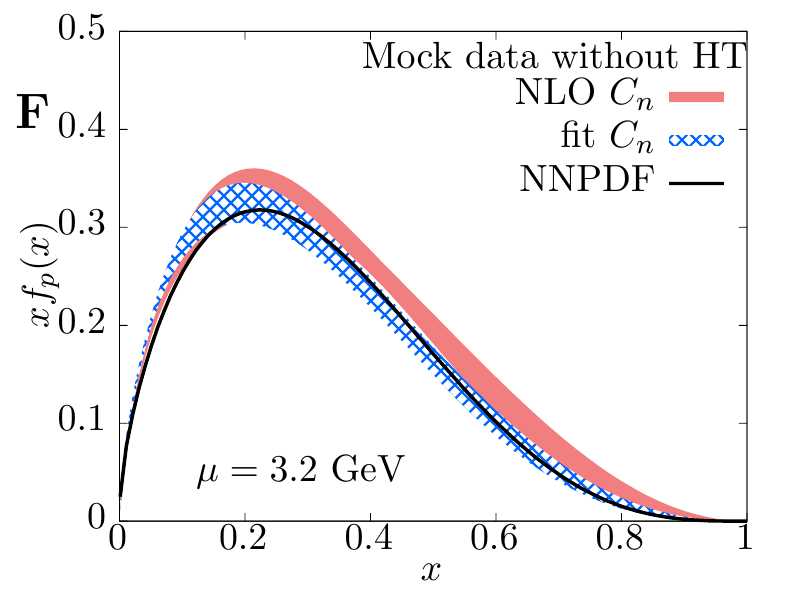}

\caption{(Top panels \textbf{A, B, C}) 
The Wilson coefficients $C_n(z^2)$ obtained from fits to the mock
pion ITD are shown. The plots \textbf{A, B, C} are for $n=2,4,6$
respectively.  The black curves are the respective NLO results for
$C_n$. By construction of the mock-pion ITD, the underlying Wilson
coefficients contain substantial corrections to the NLO $C_n$, and
hence the observed discrepancy.  (Bottom panels \textbf{D, E}) The
Mellin moments, $\langle x^2\rangle$ and $\langle x^4\rangle$,
extracted by twist-2 OPE fits to the proton pseudo-ITD over the
range of $z\in [2a, z_{\rm max}]$ are shown as a function of range
maximum $z_{\rm max}$ in the panels \textbf{D} and \textbf{E}
respectively.  The results using NLO $C_n(z^2)$ as well as the
$C_n(z^2)$ obtained from fits to the pion ITD (shown in the top
panels) are displayed together.  The corresponding NNPDF moments
that were used to produce mock-proton data are the black horizontal
lines.  In bottom panel \textbf{F}, the results from fits of the
JAM-type two parameter ansatz $f(x) = x^\alpha (1-x)^\beta$ are
shown using the two types of $C_n$.
}
\eefs{mockdemocn}

\subsection{Case-1: When the dominant corrections are higher-loop
corrections to leading-twist OPE}

First, let us look at a case where there are only corrections to
the NLO $C_n$, which we will denote as $C_n^{\rm NLO}$, via non-zero
$\Delta \tilde{c}_n(j)$ terms, and no higher-twist correction terms,
i.e., $\Lambda = 0$ for both the pion and the proton.  This is exactly the
case where the our method has to work better than the one using
finite order $C_n$, and therefore, it should not be surprising that we will
account for the corrections at the end of this subsection.
We experimented with different models of corrections to NLO Wilson
coefficient (and the reader can try their own setup in the Jupyter
notebook here~\cite{gitlink}), but we discuss an example case that
we think is more plausible, with corrections given as $\left[\Delta
\tilde{c}_n(0), \Delta \tilde{c}_n(1), \Delta \tilde{c}_n(2)\right]
= \left[0.1037, 0.0207, -0.0025\right]\ln(n+1)$.  The choice of
$\ln(n+1)$-dependence of the correction terms is inspired by the
asymptotic behavior of the Harmonic functions that appear in the
perturbative kernel and could be important to threshold resummation
as discussed in~\cite{Gao:2021hxl}.  The mock-data for ${\rm Re}{\cal
M}$ that is generated from ${\cal N}\left(\overline{{\cal M}}^0_H,
\Sigma\right)$ for the proton and the pion are shown in \fgn{mockdemo1}.

By using the JAM20~\cite{Cao:2021aci,Barry:2018ort} determination
of pion PDF as the fixed input, we fitted the coefficients
$C_n(z)$ for even values of $n>0$ as the free parameters to the
mock-pion data using the twist-2 expression in \eqn{twistope2}
with the truncation at $n=N_{\rm max}=10$.  As discussed above, for the ratio pseudo-ITD,
the value of $C_0(z)=1$, and hence it is not a fit parameter.  The
fits are analogous and non-parametric in $z$ to the type of fits
used in ``OPE without OPE" analysis~\cite{Karpie:2018zaz} of moments
as a function of $z$, with the modification being that we are now
fitting the $z$-dependent coefficients $C_n(z)$ instead of moments.
We imposed a prior on $C_n$ from $C^{\rm NLO}_n$ with a rather broad width of
$\sigma_{\rm prior}$ of 100\% of $C^{\rm NLO}_n$ at any $z$.  In
the top panels A, B and C of \fgn{mockdemocn}, we show the
three of the fitted $C_n(z)$ for $n=2, 4$ and 6 as a function of
$z$. These central (MAP) values and their errors shown in the plot (and the
covariances between different $C_n(z)$ which are not shown) summarize
the abstract posterior distribution
$Q(\{C_n\})$ we used in our discussion of the method. For comparison,
$C^{\rm NLO}_n$ is also shown as the black curve, which clearly
disagrees with the fitted $C_n$, especially for $C_2$. This is expected by construction
in this example mock-data. For $z/a=1$ for $C_4$ and $z/a \le 4$
for $C_6$, the fitted values fall exactly on the NLO
curve. This is due to the prior imposed, as there is no information on $C_n$ 
for these $z$ and its knowledge is purely derived from the prior. 
As $z$ is increased, one sees a cross-over from a prior-driven fit to a more
data-driven one.

In the second step, we used the $C_n$ obtained above in the twist-2
OPE analysis of the mock proton pseudo-ITD data shown in \fgn{mockdemo1}.
We performed a variant of the ``OPE without OPE" analysis, in which
we fitted proton moments $\langle x^2 \rangle_p, \langle x^4
\rangle_p$ and $\langle x^6 \rangle_p$ as free fit parameters in
the twist-2 OPE expression, to the proton mock-data satisfying, $z
\in [2a, z_{\rm max}]$ and all momenta. We took care of the uncertainty
in $C_n$ by randomly drawing from the bootstrap samples obtained
in the analysis of the pion in the first step. We repeated these
fits by increasing $z_{\rm max}$ gradually.  We have shown the
dependence of the fitted moments  $\langle x^2 \rangle_p$ and
$\langle x^4 \rangle_p$ on $z_{\rm max}$ in the bottom panels D and
E of \fgn{mockdemocn}, respectively.  As expected, $\langle x^2
\rangle_p$ from the fits using $C_n$ obtained from the pion agrees
with the NNPDF value, which is the underlying PDF used to generate
the proton pseudo-ITD mock-data. Similar to the case in actual
lattice computations presently, the signal for $\langle x^4 \rangle_p$
is poor at shorter $z_{\rm max}$, and albeit up to errors, the
results with fitted $C_n$ agrees better with NNPDF value than the
NLO $C_n$ as we venture to larger $z_{\rm max}$.  In the bottom
panel F of \fgn{mockdemocn}, we demonstrate how this method can
also be applied in the reconstruction of $x$-dependent PDF by using
a two-parameter ansatz of the form, $f(x) \sim x^\alpha (1-x)^\beta$;
for this fit, one simply writes the moments as a function of $\alpha$
and $\beta$, and then performs the fits using the twist-2 OPE
formula in \eqn{twistope2} with $\{C_n\}$ being from either NLO or
from the fits to the pion.  The PDF reconstruction analysis with fitted
$C_n$ reproduces the proton PDF, $f_p(x)$ to a better accuracy,
whereas the one using the NLO kernel does not, as expected in this
example. Through the above analysis of the mock-data, we have elaborated the
typical steps we expect to be involved in the implementation of the
method.

\subsection{Case-2: Possibility of suppressing higher-twist effects}

The method we have laid out still assumes that the leading twist
OPE describes the lattice data to a good accuracy.  However, given
that we only know the data up to errors, we can ask if the higher
twist corrections can effectively be absorbed into effective
Wilson coefficients determined by fits; we have to admit that this procedure
is not pristine. In order to look for this
possibility, let us consider a higher-twist correction to \eqn{twistope2}
of the form $\Lambda_\pi^2 z^2 \om^2$ for the pion pseudo-ITD and
$\Lambda_p^2 z^2 \om^2$ for the proton pseudo-ITD. Let us first
assume that such a higher-twist effect is more or less hadron mass
independent, and that $\Lambda_\pi \approx \Lambda_p = \Lambda$.
In the first step, when we do the fits of $\{C_n\}$ to, say the
pion, such a higher-twist term can effectively be absorbed into a
Wilson-coefficient given by
\beq
C_2^{\rm eff}(z) \equiv C_2^{\rm twist-2}(\mu^2 z^2)  - \frac{2\Lambda^2 z^2}{\langle x^2\rangle_\pi},
\eeq{effcn}
where $ C_2^{\rm twist-2}(\mu^2 z^2)$ is the part of the fitted
value that is the actual twist-2 Wilson coefficient.  When such an
effective $C_2^{\rm eff}(z)$ is used in the analysis of the proton
pseudo-ITD, the net higher-twist correction in proton gets reduced
to $\Lambda^2\left(1-\frac{\langle x^2\rangle_p}{\langle
x^2\rangle_\pi}\right)  z^2 \om^2$, which by putting in values for
the moments at $\mu=3.2$ GeV, can be estimates to be a 70\% reduction.
On the other hand, if the higher-twist correction term $\Lambda_p,
\Lambda_\pi$ were proportional to the proton and pion masses
respectively, there would hardly be any reduction in higher-twist
term for the proton by this process.  
Unfortunately, 
the property of the higher-twist correction in the lattice data are 
unknown.

We repeated the mock-data analysis presented in the previous
sub-section, but this time, included the same value of $\Lambda\ne
0$ in \eqn{opemock} for both the pion and the proton. We used a value
of $\Lambda\approx 30$ MeV, which was found in the pseudo-ITD in
RI-MOM scheme~\cite{Gao:2020ito}, and  even smaller in the ratio scheme.
We performed the analysis in the similar way as described in the
previous subsection by inputting the JAM20 pion-PDF to determine
$C_n$ (which are effective now due to non-zero $\Lambda$). We show the
result of the reconstructed proton PDF in \fgn{mockdemocn2}. 
The use of fitted $C_n$ brings the estimated PDF closer
to the actual proton PDF underlying the data when compared to the
result obtained by using the NLO $C_n$. But, the results still
disagree with the NNPDF curve due to the fact that the higher-twist
corrections are canceled only partially.

\bef
\centering
\includegraphics[scale=0.75]{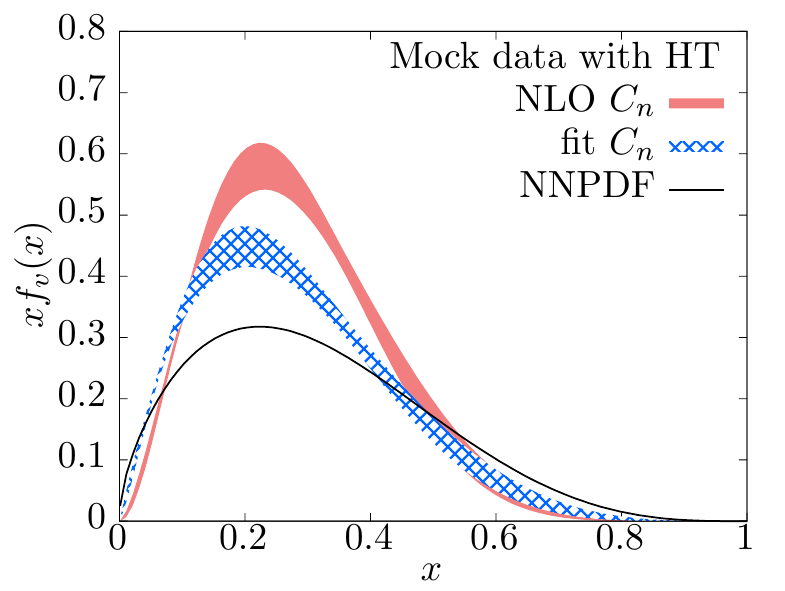}
\caption{
The result of $x$-dependent PDF fit to the proton mock-data containing
higher-twist (HT) correction (see text) using a two parameter ansatz
$f(x) = x^\alpha (1-x)^\beta$.  The results using NLO $C_n$ and
using an effective $C_n$ from fits to the mock-data for the pion
are shown. The actual underlying PDF (NNPDF) that is used to produce
the mock-proton data is shown by the black curve.
}
\eef{mockdemocn2}

\section{Application to some selected published results}

\befs
\centering
\includegraphics[scale=0.72]{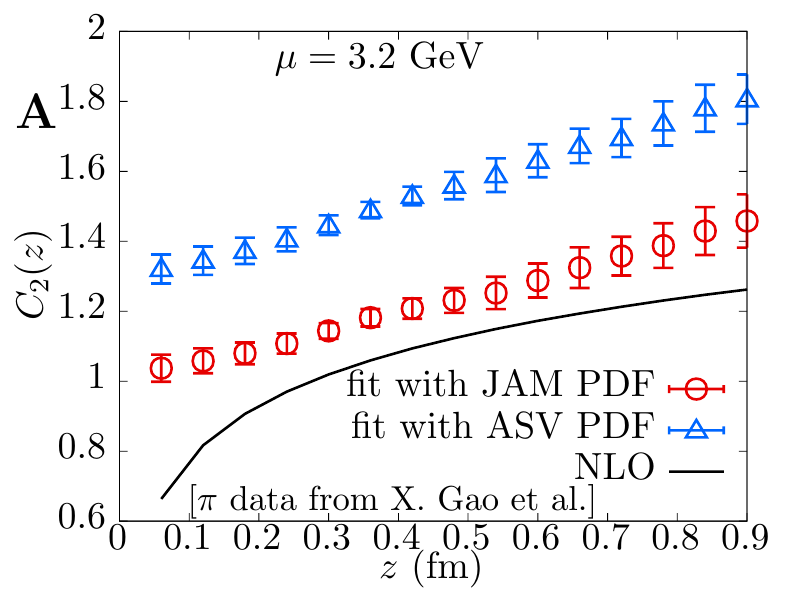}
\includegraphics[scale=0.72]{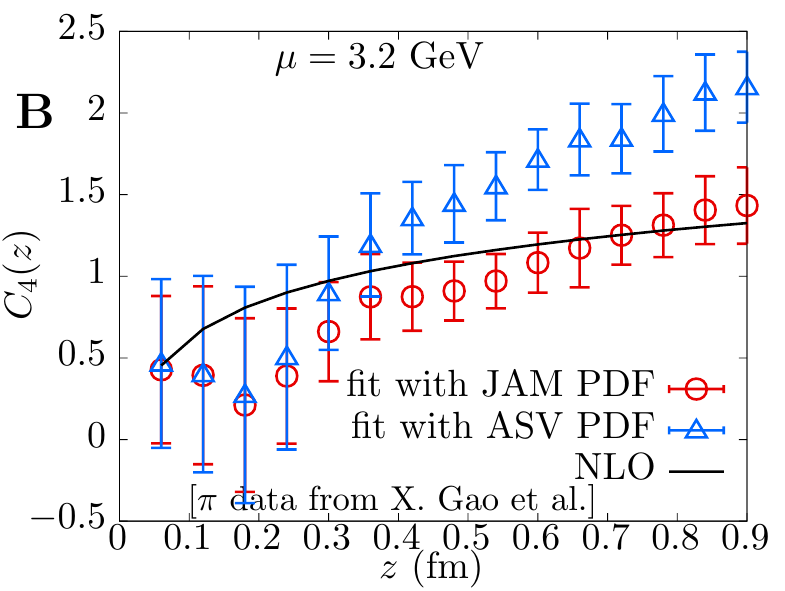}
\includegraphics[scale=0.72]{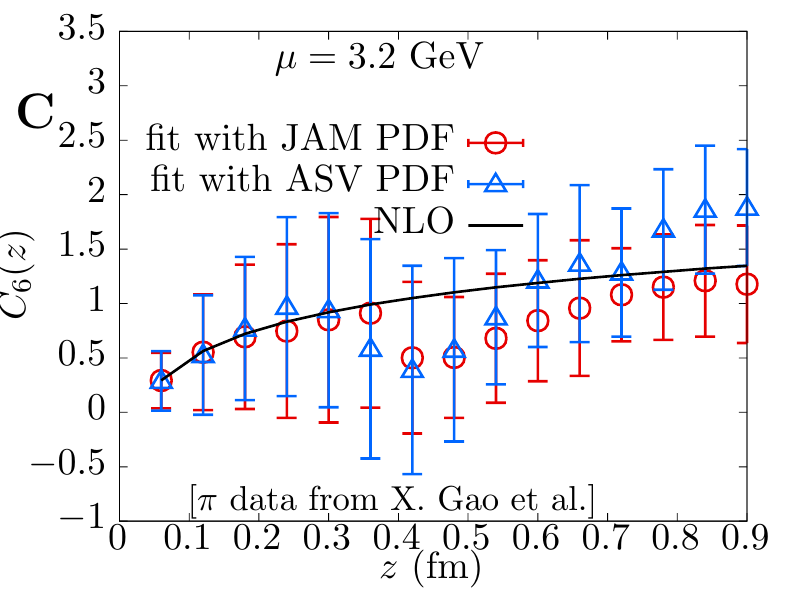}

\includegraphics[scale=0.72]{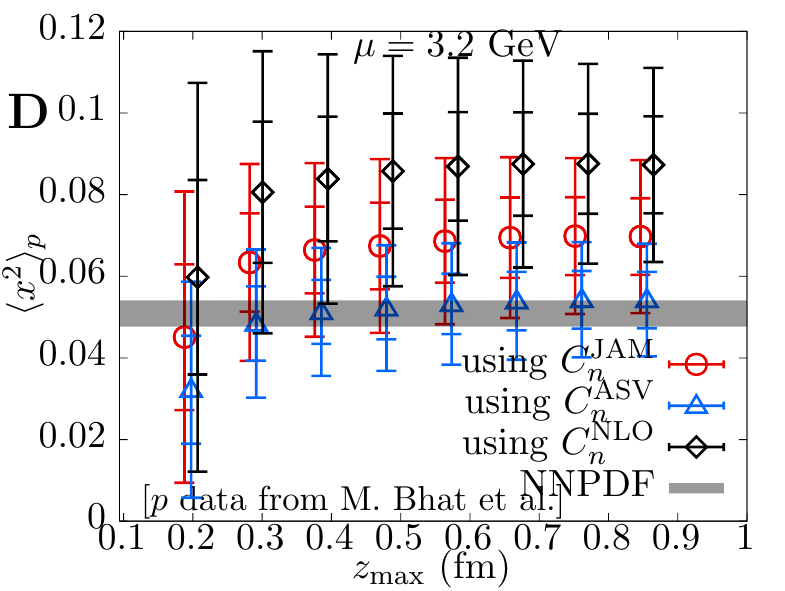}
\includegraphics[scale=0.72]{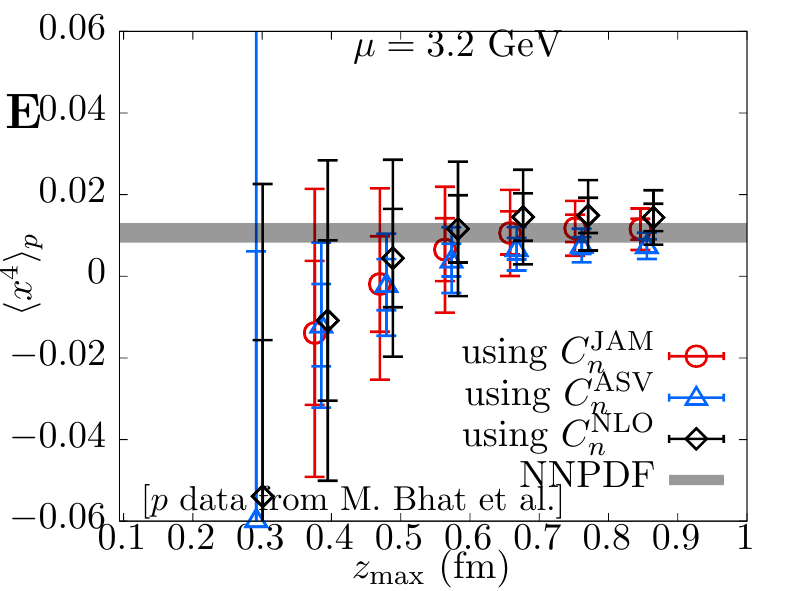}
\includegraphics[scale=0.72]{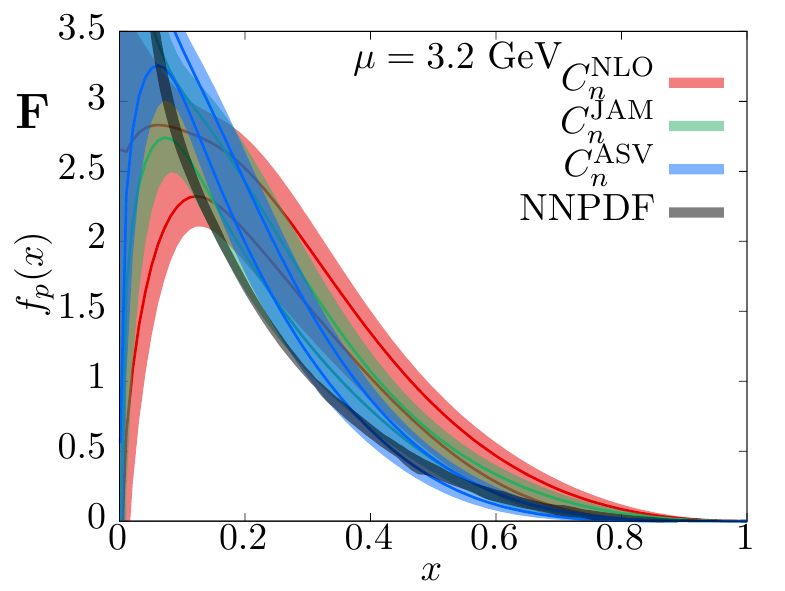}
\caption{Top panels \textbf{A, B, C} show $C_n(z)$ as a function
of $z$ obtained by fits to the pion reduced ITD data taken from a
previously published result in X.~Gao et al~\cite{Gao:2020ito}. The fits
were made using two different pheno-PDFs at $\mu=3.2$ GeV, namely,
using the JAM20 PDF (red) and the ASV soft-gluon resummed estimate
(blue). The NLO results for $C_n$ are the black curves.  Bottom
panels \textbf{D, E, F} show the application of $C_n$ extracted
from the pion, to the analysis of the proton pseudo-ITD data presented
in M.~Bhat et al~\cite{Bhat:2020ktg}. The panels \textbf{D, E} show the fitted
values of $\langle x^2 \rangle_p$ and $\langle x^4 \rangle_p$ as
function of the maximum value of $z=z_{\rm max}$ of the fit range
$z \in [2a, z_{\rm max}]$.  The red, blue and the black points
correspond to the estimates using the values of $C_n$ obtained using
JAM20 pion PDF, using ASV PDF and the NLO values respectively.  The
estimate of the moments from the NNPDF at $\mu=3.2$ GeV are shown
as horizontal black bands.  The panel \textbf{F} show the proton
PDFs reconstructed using two-parameter ansatz using the three-types
of $C_n$.
}

\eefs{latticecn}

In the last part of the paper, we apply our methodology to some of
the previously published lattice results from different collaborations. 
However, we should remark that our analysis is certainly crude as we simply take the central values
and errors presented in the publications, and we do not take into account
the correlated statistical fluctuations in the data. More importantly,
we will apply this method to some cases where the computations were
performed with heavier than physical pion masses. We believe this
is not a serious issue, at least for the pion PDF
(c.f.,~\cite{Detmold:2003tm,Sufian:2020vzb}). However, looking
forward to the future, where more computations of both the pion and
proton at the physical point might be forthcoming, we expect this
method will be more appropriate.  However, with these
cautions, let us apply the method to some of the existing data in
the literature.

The details of the published lattice QCD results for the pion and proton that we
borrow the data  (central-value and error)  from, are as follows.
For the pion, we take the ratio pseudo-ITD results~\cite{Gao:2020ito}
for a 300 MeV pion determined on an ensemble with $a=0.06$ fm and
lattice size of $48^3\times 64$.  For the proton, we take the ratio
ITD results at the physical point~\cite{Bhat:2020ktg} obtained on
an ensemble with $a=0.0938$ fm on $48^3 \times 96$ lattices.  For
complete details on the lattice methods used in these studies, we
refer the reader to the respective papers~\cite{Gao:2020ito,
Bhat:2020ktg}.

\bef
\centering
\includegraphics[scale=0.73]{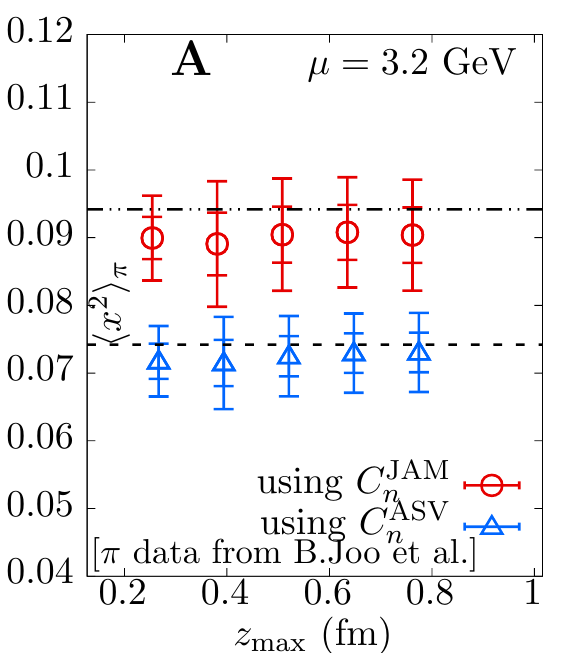}
\includegraphics[scale=0.73]{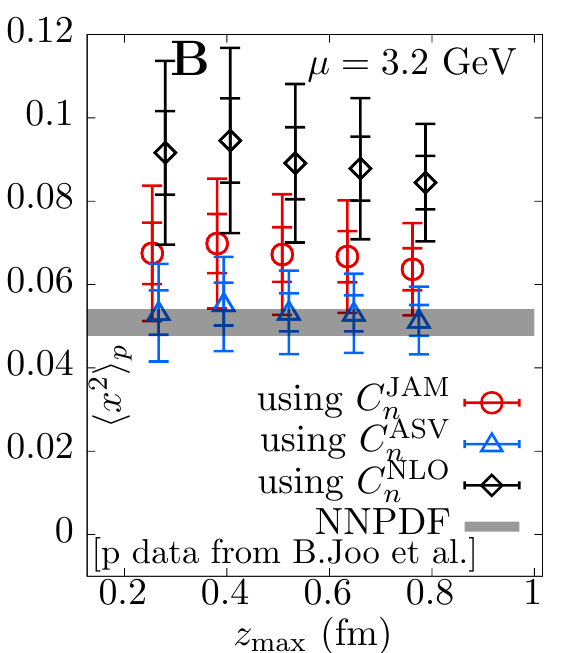}
\caption{
Panel \textbf{A}: Wilson coefficients $C_n^{\rm JAM}$ and $C_n^{\rm
ASV}$ obtained from fit to pion pseudo-ITD data (taken from X. Gao
et.al~\cite{Gao:2020ito} and shown in the top panels of \fgn{latticecn})
are applied to a completely different pion pseudo-ITD result taken
from B. Jo\'o et al.~\cite{Joo:2019bzr} to check for internal
consistency and weak dependence on mass and lattice spacing.  The
result of two sets (red and blue) of fitted moments $\langle x^2
\rangle_\pi$ obtained from the usage of $C_n^{\rm JAM}$ and $C_n^{\rm
ASV}$ are shown as a function of fit range $z\in [2a, z_{\rm max}]$.
For comparison, $\langle x^2 \rangle_\pi$ from JAM20 and ASV are
shown as dot-dashed and dashed lines.  Panel \textbf{B}: A similar
analysis of proton moment $\langle x^2 \rangle_p$ using  $C_n^{\rm
JAM}, C_n^{\rm ASV}, C_n^{\rm NLO}$  as applied to the proton data
obtained in B. Jo\'o et al.~\cite{Joo:2019jct} in the same ensemble
as~\cite{Joo:2019bzr}, and shown as a function of $z_{\rm max}$.
The NNPDF result is shown for comparison.
}
\eef{jlabx2}

We followed the same set of steps as described for the analysis of mock-data sets
for these real lattice data, and hence we will keep the details brief.
We used the pion PDF as the starting point. However, in the analysis here,
in addition to the JAM20 pion PDF, we also included the ASV pion PDF
estimate~\cite{Aicher:2010cb} (obtained by a
reanalysis of the Fermilab data~\cite{Conway:1989fs} by including soft-gluon resummation).
In the pion PDF literature, there is still an ongoing debate on
whether the large-$x$ exponent $\beta$ is closer to 1 or about 2~\cite{Nguyen:2011jy,Chen:2016sno,Bednar:2018mtf,Aicher:2010cb,RuizArriola:2002wr,Broniowski:2017wbr,deTeramond:2018ecg,Ding:2019qlr}.
In the top panels (A, B and C) of \fgn{latticecn}, we show the Wilson coefficients
$C_2$, $C_4$ and $C_6$ obtained by fits to the pion pseudo-ITD data.
The fitted values of $C_n$ obtained by assuming the JAM20 PDF as
the input PDF, are shown as red circles in the plots.  On the other
hand, when we assumed that the ASV pion PDF at the same $\mu=3.2$
GeV scale as the given input PDF, we obtained the points shown in
blue.  For brevity, we will
refer to the two sets of Wilson coefficients as $C_n^{\rm JAM}$ and
$C_n^{\rm ASV}$, respectively.  We remark that it would be a good idea to 
skip the first one or two lattice points for $C_2$ in the above plots as they are likely
to be affected by small-$z$ lattice artifacts.  To compare, the NLO
values of $C_n$ are shown as the black curves, which lie comparatively
closer to $C^{\rm JAM}_n$ than to $C^{\rm ASV}_n$.  To rephrase, 
if one is confident that the ASV result is the {\sl correct} pion
PDF, then it  implies that even the first non-trivial Wilson
coefficient $C_2$ needs to have about 40\% corrections to NLO $C_2$
resulting from higher-loop terms (given the assumption that
higher-twist terms are not as important).  For $C_2$ and $C_4$, the
fits are driven by the lattice data as well as the input PDF for
all $z$, whereas for $C_6$, the data is driven only by NLO prior
for $z< 0.3$ fm and then there is cross-over to data-driven regime for
larger $z$.

In the next step, we applied $C^{\rm JAM}_n$ and $C^{\rm ASV}_n$
determined above to the proton pseudo-ITD lattice data. We dealt with 
the mismatch of lattice spacing between the two lattice calculations by interpolation. 
We first fitted the even moments, $\langle x^2\rangle_p$ and
$\langle x^4\rangle_p$ to the proton pseudo-ITD data using twist-2 OPE truncated
at ${\cal O}(\om^4)$ containing the choices of Wilson coefficients as 
$C^{\rm JAM}_n, C^{\rm ASV}_n$ and $C^{\rm NLO}_n$ at $\mu=3.2$ GeV.  
We performed these fits including the
data with $z\in[2a, z_{\rm max}]$ for $z_{\rm max}$ up to 0.75 fm.
The results for $\langle x^2\rangle_p$ and  $\langle x^4\rangle_p$
from these fits are shown as a function of the maximum of the fit range,
$z_{\rm max}$, in the panels D and E of
\fgn{latticecn}. The black, blue and red points are the results
from such fits using $C_n^{\rm NLO}, C_n^{\rm ASV}, C_n^{\rm JAM}$
respectively. For comparison, the black bands are NNPDF estimates
of the proton moments at $\mu=3.2$ GeV. For $z_{\rm max}\approx 0.24$ fm,
there is a discernible signal for $\langle x^2\rangle_p$,
while $\langle x^4 \rangle_p$ remains comparatively noisier. To
account for the possibility of an underestimation of the errors in our analysis
due to the absence of any correlated statistical fluctuations,
we conservatively included both the 1-$\sigma$ as well as the 2-$\sigma$ errors
in the plots for the moments.  The NLO result for
$\langle x^2 \rangle_p$ is a bit higher than the NNPDF
value (as also observed in the actual full-fledged analysis in~\cite{Bhat:2020ktg}). 
The usage of $C_n^{\rm JAM}$ leads to a slight decrease in
the estimated value of $\langle x^2 \rangle_p$, and surprisingly,
the largest change comes from the usage of $C_n^{\rm ASV}$, which
lowers the estimated $\langle x^2 \rangle_p$ that also agrees better with the NNPDF
value.  For $\langle x^4 \rangle_p$, there are corresponding changes
in the central values depending which $C_n$ is used, but completely masked by the larger error-bars.
In the bottom panel-F, we show the unpolarized proton
PDF, $f_p(x)$, that we reconstructed using a simple, $f_p(x)\sim x^\alpha
(1-x)^\beta$ ansatz. Again, the results obtained using the three types of
$C_n$ are shown in the different  colors; the 2-$\sigma$ estimate
is the outer-band and 1-$\sigma$ result is the inner-band enclosed
by the solid lines. The $x$-dependent results are
sensitive to which Wilson coefficient is used, and particularly,
the result obtained using $C_n^{\rm ASV}$ agrees better with the
NNPDF curve.

In order to see if the results presented above are some artifacts of putting
together the results from two completely different lattice calculations,
at different pion masses (with only the proton at physical point)
and at different lattice spacings, we repeated the above analysis
to the central-value and error data of the pion and proton pseudo-ITDs
that we borrowed from the papers~\cite{Joo:2019bzr, Joo:2019jct}
by the HadStruc collaboration. The results reported in these papers
used $32^3\times 96$ lattice with $a=0.127$ fm at a heavier pion
mass of 415 MeV. The analysis of the 415 MeV pion data from
Ref.~\cite{Joo:2019bzr} by applying the set of $C_n$ obtained also
from the pion, with a 300 MeV mass and from a completely different
data in~\cite{Gao:2020ito}, is an interesting exercise. If there
is not much lattice spacing and pion mass dependence in the two
pion data, then the analysis using $C_n^{\rm JAM}$ from panels-(A,B,C)
in \fgn{latticecn} should result in the JAM20 moments as it is the
initial input PDF in this case, and similarly, an analysis using
$C_n^{\rm ASV}$ should result in ASV moments.  In the panel-A of
\fgn{jlabx2}, we show $\langle x^2\rangle_\pi$ from such an
analysis by fitting the moments over a range $z\in [2a, z_{\rm
max}]$.  We have again conservatively plotted both the 1-$\sigma$
and the 2-$\sigma$ errors.  It is satisfying that this pion analysis
reproduces the JAM20 and ASV moments as per our initial expectation.
In the right panel-B of \fgn{jlabx2}, we similarly show the result
of applying $C_n^{\rm JAM}, C_n^{\rm ASV}$ and $C_n^{\rm NLO}$ to
the proton data from~\cite{Joo:2019jct}. The result is very similar
to the one we observed in the panels-(D,E) of \fgn{latticecn}, with
a similar effect of the choice of $C_n$ on the determined moments.

Given the cautions we offered at the beginning of this section, we
need to be careful about the conclusions we can draw from the
observations we have made. First, one should pay attention to fact
that the usage of ASV versus JAM20 pion PDF as input PDFs leads to
changes in $C_n(z)$ even for $n$ as small as 2.  To contrast, in
Ref.~\cite{Gao:2021hxl}, the corresponding threshold resummation
for the pseudo-ITD calculation was investigated as the effect of
potentially large $\alpha_s \ln(n)$ and $\alpha_s\ln^2(n)$ terms
that will be present in $C_n$ at larger $n$, and thus will not
affect the smaller $n$.  Thus it is paradoxical that imposing the
correctness of ASV PDF automatically implies a big correction to
$C_n$ even for small $n$; we do not have a resolution in this paper.
Thus, it would be important to repeat the exercise presented in
this section to the lattice data for pion pseudo-ITD at the physical point 
when available in the future to see if the observation we have made still stands.
Secondly, we see that an unaccounted larger higher-loop correction
to $C_n$ starting from $n=2$ is sufficient  for a better
agreement between lattice computations of proton PDF 
and  the phenomenological estimates, without any necessity to invoke
higher-twist contaminations to the leading-twist OPE.  The curious
agreement of the lattice estimation for the 
proton moments and the PDF when $C_n^{\rm ASV}$ is used in the analysis instead of NLO
$C_n$, brings forward the great impact a better determination of
the pion PDF from global analysis in the future (c.f.,~\cite{Aguilar:2019teb,Denisov:2018unj}) will have on a
lattice computations of proton PDFs; a somewhat unexpected and
interesting connection due to the method presented here.

\section{Discussion}

The method we advocated in this paper for lattice determination of valence 
PDFs consists of the following steps: (1)
The leading-twist Wilson  coefficients $C_n(\mu z)$ are determined
directly by fitting the lattice data for a hadron, say the pion,
by assuming the pion PDF is known well from phenomenological
determinations and then taken as an input for the lattice analysis.
At this step, our degree of confidence that a perturbative calculation
at a given order is sufficient can be incorporated via addition
of priors to the fits. (2)  The $C_n(z)$ obtained from fits is then
used in the analysis of lattice data to determine the PDF of another
hadron, say, the proton by performing the usual reconstruction analysis
methods to determine the $x$-dependent PDF, except that one replaces
the perturbative matching coefficients with the one obtained from
the pion, in this example. (3) One can also use the fitted values
of $C_n$ in the analysis of zero-skewness GPD of, say the pion
itself, at non-zero momentum transfers.  We demonstrated the
feasibility of this method by applying it to a set of mock-data, as well as to
some published lattice pseudo-ITD data for the pion (at 300 MeV and 415 MeV)
and the proton (at physical point and at 415 MeV). In doing so, we showed how 
the precise extraction of pion-PDF from the experimental data can have 
an impact on lattice studies of proton PDF, indirectly via the method 
we presented.

A drawback of the method is that there should be a good determination
of pheno-PDF; this is achievable for the valence unpolarized PDF, but fails
for the nonsinglet helicity and transversity PDFs since they are defined only
for the proton. In such cases, the analysis strategy (3) is still
viable, i.e., extend to zero-skewness helicity and transversity GPD
of proton by inputting the experimental values of helicity and
transversity PDF of proton respectively. This can also be tested 
in the helicity GPD data already presented in~\cite{Alexandrou:2020zbe}.
In the future, it would also be interesting to extend this study 
to the iso-singlet PDF cases wherein the small-$x$ uncertainties
in the global fit PDF themselves might need to be propagated into the analysis of lattice
matrix elements (for example, see Ref.~\cite{Ball:2017otu} for a discussion of 
uncertainties from small-$x$ logarithms in global fit determination of singlet PDFs).

One could ask why should one perform this rather complicated procedure
when one is, say,  75\% certain that the correction to the Wilson
coefficient $C_2(z)$ from higher-loop calculation will be less than
2\%.  In fact, the point of the paper was to generalize the
perturbative matching method in such a way as to fold in our
uncertainties about perturbative Wilson coefficients itself in the
analysis; the Bayesian approach achieves this for us.  In the
example, we would arrange the Bayesian prior distribution of $C_2(z)$
in such a way that it is 75\% likely that the $C_2$ lies within
$2\%$ of the NLO Wilson coefficient. This way, we are making the
degree of our ignorance of $C_n$ quantitative.

A reasonable concern could be that the experimental data is known
only up to NLO or NNLO order, and hence, why should one use a procedure
other than using NLO or NNLO value of $C_n$ for matching. The assumption
we are implicitly making here is that the perturbative convergence
of pheno-PDFs is fast and well-understood, whereas it does not
necessarily guarantee a rapidly converging perturbative Wilson
coefficients that are used to match PDFs to the lattice matrix
elements.  Thus, the solution, which in retrospect appears very simple,
is to replace the two different perturbative calculations (one to
extract pheno-PDF and the other to match lattice results), with a
single well-studied perturbative step that has been used to determine
the pheno-PDF via global analysis.

\section*{Acknowledgments}

We thank Martha Constantinou and Krzysztof Cichy for sharing with us their
nucleon ITD data points. We thank the BNL-based collaboration for
sharing their data points for the pion ITD. We thank the HadStruc
collaboration for sharing their pion-ITD and nucleon-ITD data points.
We thank Patrick Barry for helping us with JAM20 pion PDF data sets.  We
are thankful for the insightful discussions with Jianwei Qiu which
assisted our work. We thank Joe Karpie, Swagato Mukherjee, Kostas
Orginos, Peter Petreczky and David Richards for their comments on
the manuscript. We thank Yong Zhao for clarifications regarding
the GPD matching. The authors are supported by U.S. DOE grant No.
DE-FG02-04ER41302 and Jefferson Science Associates, LLC under U.S.
DOE Contract No. DE-AC05-06OR23177

\bibliography{pap.bib}

\end{document}